# Analysis of Scanned Probe Images for Magnetic Focusing in Graphene

Sagar Bhandari[1], Gil-Ho Lee[2], Philip Kim[1,2] and Robert M. Westervelt[1,2]

1 School of Engineering and Applied Sciences, Harvard University

Cambridge, MA 02138, U.S.A

2 Department of Physics, Harvard University

Cambridge, MA 02138, U.S.A

sbhandar@fas.harvard.edu

**Abstract**

We use a cooled Scanning Probe Microscope (SPM) to electron motion in nanoscale devices. The charged tip of the SPM is raster scanned at a constant height above the surface as the conductance of the device is measured. The image charge scatters electrons away, changing the path of electrons through the sample.[1-3] Using this technique, we have imaged cyclotron orbits[3] for ballistic hBN-graphene-hBN devices that flow between two narrow contacts in the magnetic focusing regime. Here we present an analysis of our magnetic focusing imaging results based on the effects of the tip-created charge density dip on the motion of ballistic electrons. The density dip locally reduces the Fermi energy, creating a force that pushes electrons away from the tip. When the tip is above the cyclotron orbit, electrons are deflected away from the receiving contact, creating an image by reducing the transmission between contacts. The data and our analysis suggest that graphene edge is rather rough, and electrons scattering off the edge bounce in random directions. However, when the tip is close to the edge it can enhance transmission by bouncing electrons away from the edge, toward the receiving contact. Our results demonstrate that a cooled SPM is a promising tool to investigate the motion of electrons in ballistic graphene devices.

**Introduction**

Scanning probe microscopy (SPM) has been used in imaging electron motion in two dimensional electron gas (2DEG) inside a GaAs/AlGaAs heterostructures [1,2] and electronic states in a quantum dot [3,4,5]. Recently, we used SPM to image ballistic electron motion in graphene under a perpendicular magnetic field. The electrons follow cyclotron trajectories and regions in graphene corresponding to these cyclotron orbits were observed [6,7]. The sample is a hBN-graphene-hBN device etched into a hall bar geometry with two narrow (*700 nm*) contacts along each side, separated by *2.0 μm* and large source and drain contacts at either end. The heavily doped Si substrate acts as a back-gate, covered by a *285 nm* insulating layer of *$SiO_2$*. The degree of magnetic focusing is measured by injecting current $I_s$ into a narrow contact and measuring the voltage developed $V_i$ at the second narrow contact on the same side of the hall bar device. The trans-resistance $R_m = V_i/I_s$. We use a 20 nm wide conducting tip as a local probe which is raster scanned at a constant height above the sample surface as the change in conductance across the sample is measured. The image charge created by the tip on the sample surface scatters electrons away, thereby revealing the path of electrons through the sample.

In this paper, we present the theoretical model used for simulating the electron trajectories in graphene. The model is purely classical and the simulation results obtained agree quite well with the SPM data. Using these ray tracing simulations, we study the effects of the tip-created local charge density dip on the motion of electrons in graphene. We also study the regions with dip (red) or increase in trans-resistance (blue) in the SPM images. We find that the scattering of electrons from the graphene edge is diffusive. However, when a tip is placed close to the edge, the scattering of electrons is specular due to the smoothly varying charge density profile created by the tip on the sample surface. As a result, there's an increase in the number of electrons reaching the receiving contact.

**Method**

We use classical ray tracing to simulate electron trajectories in graphene under an applied magnetic field. Each electron is modelled as a particle in the two-dimensional (2D) space of the sample with time-dependent position and velocity. External forces acting on the electrons are – (1) Lorentz force due to the perpendicular magnetic field *B* and (2) force from the tip-induced charge density profile. Depending on these forces, the electrons accelerate or decelerate. The trajectories are injected from a source with cosine angular distribution that follows from the foreshortening of the apparent contact width. Fig. 1(a) shows the distribution of the injected electron trajectories from a narrow source contact without a magnetic field. The color represents the intensity of the trajectories spatially with bright yellow as most densely populated region of such trajectories and black as the least populated.

For the simulations, the source is *700 nm* wide with the injectors (point sources) uniformly spaced at an interval of 50 nm. The total number of electron trajectories $N = 10,000$. Fig 1(b) shows these trajectories after a perpendicular magnetic field B is introduced. Classically, when a magnetic field is applied perpendicular to the plane of two dimensional electron gas, the electrons travel in cyclotron orbits. The equation of motion is governed by the Lorentz force.

$$\vec{F} = e\,\vec{v} \cdot \vec{B} \tag{1}$$

where, $e$ is the elementary charge, $v$ is the electron velocity and $B$ the applied magnetic field. The radius of these cyclotron orbits is given by the following equation.

$$r_c = m^* v / eB \tag{2}$$

where $m^*$ is the dynamical mass of the electron. The conical band structure of graphene yields a linear dispersion relation $E = \hbar v_F k$ at energies close to the Dirac point, where the speed $v_F \sim 10^6 \, m/s$ is fixed [8,9,10]. Unlike conventional semiconductors, the dynamical mass is density dependent $m^* = \hbar(\pi n)^{1/2}/v_F$, and the cyclotron radius $r_c = m^* v_F / eB$. When the cyclotron orbit diameter $2r_c$ equals to the spacing $L$ between the source and drain in the graphene device, the first magnetic focusing peak in transmission between contacts occurs. The focusing field $B_F$ is given by

$$B_F = \hbar v_F (\pi n)^{1/2} / peL \quad (3)$$

where $p$ is an integer.

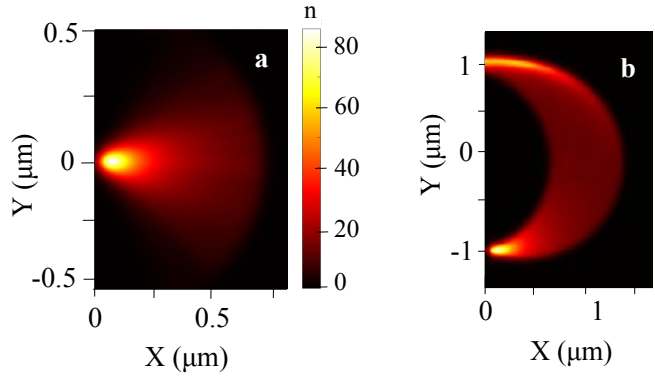

**Fig. 1:** Electron flow from a point contact into a graphene sheet (a) under zero magnetic field and (b) at the first magnetic focusing field $B_f$. (a) Distribution of electron trajectories from a point contact. These trajectories are modeled as classical particles with number of injected particles proportional to the cosine of the angle to the normal to the source. (b) Electron trajectories under magnetic field flowing between two point contacts on the first magnetic focusing peak $B = B_f$ when the cyclotron diameter $2r_c$ equals the contact spacing [6].

The electrons feel a second force created by the tip, which is positioned at a fixed height ~70 nm above the graphene surface. The difference in work function between graphene and Si tip creates an image charge density profile in the graphene sheet.

$$e\Delta n_{tip}(a) = -qh/2\pi(h^2 + a^2)^{3/2} \quad (4)$$

where $a$ is the radial distance from the tip location, $h$ is the height of the tip from the graphene sample, and $q$ is the charge on the tip. Figure 2(a) shows the charge density profile with $h$ = 70 nm, $a$ = 0, and $\Delta n_{tip}(0) = 6 \times 10^{11} cm^{-2}$, values chosen to match the data, and Fig. 2(b) shows a line cut through the center of the charge density profile in Fig. 2(a).

The image charge density profile created by the tip results in an effective force on electrons in the graphene sample. The density reduction $\Delta n_{tip}(a)$ locally reduces the Fermi energy $E_F(n + \Delta n_{tip})$. The total chemical potential $U(a) + E_F(a)$ remains constant in space, where $U(a)$ is the potential energy profile created by the tip. Therefore, the force generated by the tip is $\vec{F}_{tip}(a) = -\overrightarrow{\nabla U}(a) = \overrightarrow{\nabla E_F}(a)$. This yields the following equation of motion for the electrons passing near the tip:

$$d^2\vec{r}/dt^2 = (1/2)v_F^2 \overrightarrow{\nabla n}(\vec{r})/n \quad (5)$$

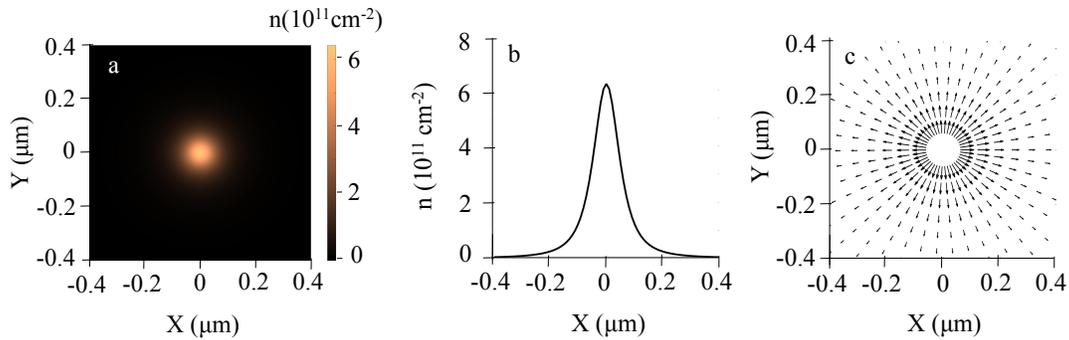

**Fig. 2:** Tip induced (a) charge density profile and (c) resulting force profile. Fig (a) Density change of the image charge profile in the 2D graphene sheet created by the tip. (b) Line cut through the center of the charge density profile. (c) Force profile felt by the electrons in graphene associated with the charge density profile.

Electrons are pushed away from the area of low density underneath the tip. Figure 2(c) shows the force profile felt by the particle in the plane. The force points radially outward with maximum at the center and gradually decreases in strength with distance from the tip center.

The ray tracing simulations are based on the classical motion of electrons through the device, subject to the magnetic field and tip perturbation. The transmission $T$ of electrons between point contacts 1 and 2 is computed by counting the fraction of emitted trajectories that reach the receiving contact. When present, the tip scatters $N$ electron trajectories away from the receiving contact, changing the number received from $p_i$ to $p_{tip}$ and the transmission by $\Delta T = (p_i - p_{tip})/N$. Because electrons hitting the receiving contact do not pass into the voltmeter, the local density and chemical potential build up to create an opposing current that nulls the total electron flow. In the experiments, the transmission change $\Delta T$ is measured by the voltage change $\Delta V_s$ of the receiving contact, and the corresponding transresistance change $\Delta R_m = V_s/I_i$ where $I_i$ is the injected current at the first contact.

**Results**

Figure 3(a) shows a cooled scanning probe image of cyclotron orbits for electrons in graphene at 4.2 K, showing the trans-resistance change $\Delta R_m$ vs. tip position at $B = 0.12\ T$ for electron density $n = 1.13 \times 10^{12} cm^{-2}$. We chose a magnetic field just above the first magnetic focusing peak $B_f = 0.11\ T$ at this density. The semicircular red region is an image of cyclotron orbits flowing between the two contacts; the transmission is decreased along the cyclotron orbits, because the tip scatters electrons away from the receiving contact. In addition, a blue region is shown near the sample edge, where the tip enhances transmission by preventing electron orbits from being scattered by the edge of the sample. Fig. 3(b) shows ray tracing simulations of the change in transmission $\Delta T$ caused by the tip vs. tip position at $B = 0.14\ T$ and $n = 1.13 \times 10^{12} cm^{-2}$. The simulations show features that are quite similar to the experimental results: an decrease in transmission along the cyclotron orbit that connects the two contacts, and enhanced transmission, when the tip bounces electrons away from the diffusely scattering edge of the sample.

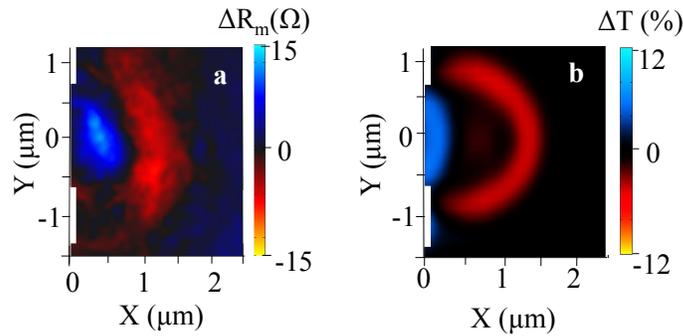

**Fig. 3:** (a) Cooled scanning probe microscope image of cyclotron orbits between two narrow contacts (white bars on left) mapping the change in transresistance $\Delta R_m$ at $B = 0.12$ T and density $n = 1.13 \times 10^{12}$ cm$^{-2}$. (b) Ray tracing simulations for the electron transmission map at B =0.14 T and n = 1.13 x 10$^{12}$ cm$^{-2}$ [6].

**Discussion**

As shown in Fig. 4(a), the tip deflects the electrons away from the receiving contact and thereby reducing the number of trajectories reaching the contact. In this figure, at a fixed tip position $(x, y) = (1.2, 0)\ \mu m$, the electron trajectories are plotted from the simulation. The grey lines represent the trajectories that reach from source to the receiving contact – these trajectories are unperturbed by the tip. The red lines represent the trajectories that get deflected by the tip away from the receiving contact. This lowers the voltage at the receiving end and therefore, trans-resistance drops. Therefore, the red region show the path of electrons present in the sample.

Figure 4(b) shows the electron trajectories when the tip is positioned at $X = 0.15$ µm, $Y = 0$ µm. The blue electron trajectories are deflected by the tip into the receiving contact, increasing the transmission between contacts. The grey trajectories fail to make it to receiving contact. When the tip is not present, the electron trajectories hitting the edge would bounce in random directions due to roughness. The edge in the hBN/graphene/hBN sandwich is patterned by a Reactive Ion Etcher (RIE) into the Hall bar geometry.

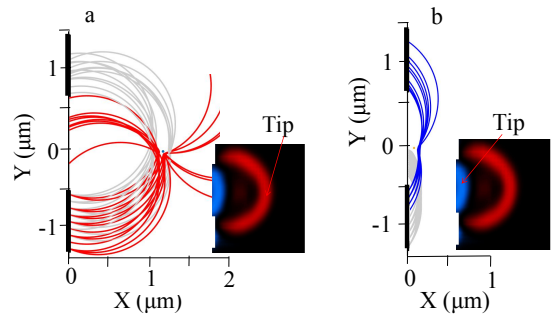

**Fig. 4:** (a) Ray-tracing trajectories for $B = 0.14$ T and $n = 1.13 \times 10^{12}$ cm$^{-2}$ and tip position $X = 1.2$ µm, $Y = 0$ µm, as shown in the simulated image inset. The red rays decrease the transmission when the tip deflects electrons away from from the receiving contact. (b) Ray-tracing trajectories at tip position $X = 0.15$ µm, $Y = 0$ µm, as shown on the simulated image inset. The blue regions increase the transmission when the tip deflects electrons away from the diffusly scattering edge into the receiving contact.

To show the origin of the tip enhanced transmission between contacts, Fig. 5(a) shows ray-tracing simulations of electron trajectories without the tip present for $B = 0.14\ T$ and d $n = 1.13 \times 10^{12} cm^{-2}$, the same condition as for Fig. 4(b). We simulate the reflection of the trajectories from the rough edge as a diffusively scattering model, in which the reflection angle of an incident electron is random, multiplied by the cosine of the angle of reflection associated with the apparent width of the contact. The majority of trajectories are reflected normal to the reflecting surface and very few get reflected parallel to the surface. Figure 5(b) shows ray-tracing trajectories of a beam of electrons incident on the edge in zero magnetic field (green). The trajectories are diffusively scattered as shown by the black lines. The probability distribution function of these lines is $P(\theta) = (1/2)cos\theta$, where $\theta$ is the angle of reflection.

Maxwell's equations create an integrable divergence in the charge density of the graphene sheet at its edge. The graphene device is electrically gated by a conductive Si substrate below an insulating 280 nm thick oxide. The graphene sheet can be modelled as a semi-infinite parallel plate atop an infinite conducting plane. The electric field lines between these planes can be derived using conformal mapping technique [11]. At a distance δx near the edge of the graphene sheet at potential $V_o$ to ground, the electric field is:

$$E \approx V_o/(2\pi a\delta x)^{1/2} \qquad (6)$$

where *a* is the spacing between graphene and Si substrate. The electric field and the surface charge density are inversely proportional to the square root of the distance from the edge. Because the electron density *n* is diverging near the edge, the dynamic mass $m^*$ and the cyclotron radius $r_c$ increase correspondingly. In our simulations, we haven't taken this into account. It would be important to include these edge effects in future computations to get a more accurate view of the experimental data.

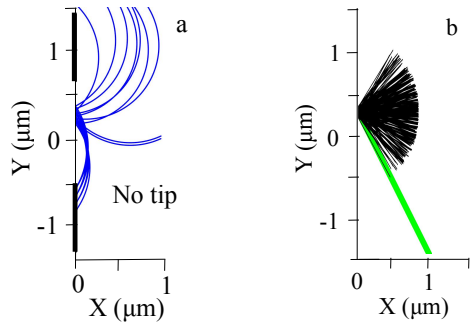

**Fig. 5:** (a) Ray-tracing trajectories for B = 0.14 T and n = 1.13 x $10^{12}$ cm$^{-2}$ with no tip present. Trajectories that would produce a second magnetic focusing peak by bouncing off the edge, are diffusively scattered by edge roughness. (b) Zero magnetic field ray-tracing trajectories of electrons incident on the edge (green) diffusively scattered (black).

**Conclusion**

The ray-tracing method accurately describes images of cyclotron orbits in graphene taken by a cooled scanning probe microscope in the ballistic regime. Using classical equations of motion with two forces: 1) Lorentz force and 2) The force due the image charge density profile created by the tip, the simulations provide a good match to the SPM data. The force due to the image charge density profile created by the tip forces electrons away from the tip location. The tip can image cyclotron orbits connecting two point contacts, by deflecting electrons away from the second contact. The tip can also enhance conductance between contacts by deflecting trajectories away from the rough edge into the second contact. The ray tracing technique could be further improved by adding the fringing electric field along the edges of the graphene device.

**Acknowledgements**

The SPM imaging research and the ray-tracing simulations were supported by the U.S. DOE Office of Basic Energy Sciences, Materials Sciences and Engineering Division, under grant DE-FG02-07ER46422. Graphene sample fabrication was supported by Air Force Office of Scientific Research contract FA9550-13-1-0211. Nanofabrication was performed in the Center for Nanoscale Systems (CNS) at Harvard University, a member of the National Nanotechnology Coordinated Infrastructure Network (NNCI), which is supported by the National Science Foundation under NSF award ECCS- 1541959.

**References**

[1] M.A. Topinka, B.J. Leroy, S.E.J Shaw, E.J. Heller, R.M. Westervelt, K.D. Maranwoski, A.C. Gossard, Science 289, 2323 (2000)

[2] K.E. Aidala, R.E. Parrott, T. Kramer, E.J. Heller, R.M. Westervelt, M.P. Hanson, A.C. Gossard, Nat. Phys. 3, 464 (2007)

[3] P. Fallahi, A.C. Bleszynski, R.M. Westervelt, J. Huang, J.D. Walls, E.J. Heller, M. Hanson, A.C. Gossard, 2005 Nano Lett. 5, 223 (2005)

[4] A.C. Bleszynski-Jayich, F.A. Zwanenburg, R.M. Westervelt, A.L. Roest, E.P.A.M Bakkers, L.P. Kouwenhoven, Nano Lett. 7, 295 (2007)

[5] A.C. Bleszynski-Jayich, L.E. Froberg, M.T. Bjork, H.J. Trodahl, L. Samuelson, R.M. Westervelt, Phys. Rev. B

77, 245327(2008)

[6] S. Bhandari, G.H. Lee, A. Klales, K. Watanabe, T. Taniguchi, E. Heller, P. Kim, R.M. Westervelt, Nano Lett. 16, 1690 (2016)

[7] S. Bhandari, Imaging electron motion in graphene, (PhD thesis, 2015)

[8] P.R. Wallace, Phys. Rev. 71, 622 (1947)

[9] A.H. Castro Neto, F. Guinea, N.M.R Peres, K.S. Novoselov, A.K. Geim, Rev. Mod. Phys. 81, 109 (2009)

[10] A.K. Geim, K.S. Novoselov, Nat. Mat. 6, 183 (2007)

[11] M.L. Boas, Mathematical Methods in the Physical Sciences, 3rd edn. (Wiley, 2006), pp. 714-716